\documentclass[aps,prb,reprint,amsmath,amsfonts,amssymb,superscriptaddress]{revtex4-2}
\usepackage{graphicx,calc}
\usepackage{color,bm}
\usepackage[normalem]{ulem}
\usepackage{braket}
\usepackage[colorlinks,urlcolor=blue,citecolor=blue,linkcolor=blue]{hyperref}
\usepackage[dvipsnames]{xcolor}
\usepackage{ulem}

\begin{document}
\title{Probing phase transitions in non-Hermitian systems with quantum entanglement}

\author{Ling-Feng Zhang}
\affiliation{Department of Physics, City University of Hong Kong, Kowloon, Hong Kong, China}
\author{Wing Chi Yu}\email{wingcyu@cityu.edu.hk}
\affiliation{Department of Physics, City University of Hong Kong, Kowloon, Hong Kong, China}

\date{\today}

\begin{abstract}
We study the quantum entanglement and quantum phase transition of the non-Hermitian anisotropic spin-$\frac{1}{2}$ XY model and XXZ model with the staggered imaginary field by analytical methods and numerical exact diagonalization, respectively. Various entanglement measures, including concurrence, negativity, mutual information, and quantum coherence, and both biorthogonal and self-normal quantities are investigated. Both the biorthogonal and self-normal entanglement quantities, except the biorthogonal concurrence, are found to be capable of detecting the first-order and $\mathcal{PT}$ transitions in the XXZ model, as well as the Ising and $\mathcal{RT}$ transitions in the XY model. In addition, we introduce the unconstrained concurrence and demonstrate its effectiveness in detecting these transitions. On the other hand, the Berezinskii-Kosterlitz-Thouless (BKT) transition in the XXZ model is revealed through concurrence and negativity at small non-Hermiticity strengths. Notably, the critical points observed in the Hermitian limit evolve into exceptional points as the strength of the non-Hermiticity increases. Furthermore, we find that the first-order transition survives up to a higher non-Hermiticity strength compared to the BKT transition within the $\mathcal{PT}$-symmetric regime of the XXZ model. 
\end{abstract}

\date{\today}
\maketitle

\section{Introduction}
Quantum phase transitions (QPTs) occur at zero temperature when the ground state of many-body systems undergoes abrupt changes due to its quantum fluctuations \cite{Sachdev_2000}. Understanding the phases and the related phase transitions has been one of the central topics in condensed matter physics. In general, there are mainly two kinds of QPTs: traditional QPTs governed by the Landau-Ginzburg-Wilson paradigm, which are conventionally characterized through the framework of local order parameters \cite{Sachdev_2000}, and those that cannot be described in this framework, such as topological QPTs \cite{RevModPhys.82.3045,RevModPhys.83.1057} and Berezinskii-Kosterlitz-Thouless (BKT) phase transitions \cite{Berezinsky:1970fr,Kosterlitz:1973xp}. Many elusive phases emerge as the interaction comes into play, where order parameters are hard to find or cannot be described by the order parameters. Since pioneering works investigated the relation between quantum entanglement and QPTs \cite{Osterloh2002,PhysRevA.66.032110,PhysRevLett.90.227902}, quantum information concepts have been successfully used in the study of QPTs. These concepts include entanglement \cite{Osterloh2002,PhysRevLett.93.086402}, mutual information \cite{PhysRevA.82.022319}, quantum coherence, quantum fidelity \cite{PhysRevE.74.031123,PhysRevLett.96.140604}, fidelity susceptibility \cite{PhysRevE.76.022101} and quantum discord \cite{PhysRevLett.88.017901,PhysRevB.78.224413,PhysRevA.80.022108}.  The lack of a priori knowledge of the order parameter and the symmetry of the system makes this method advantageous, thus it has been an important part in characterizing and understanding QPTs \cite{Osterloh2002,Sachdev_2000,PhysRevA.66.032110}. 

On the other hand, great attention has been paid to non-Hermitian physics due to the rapid experimental progress in recent years \cite{moiseyev2011non,RevModPhys.93.015005,Ashida02072020}. Non-Hermitian systems can be readily realized in multiple experimental platforms ranging from photonics \cite{Xiao2020} and phononics \cite{PhysRevLett.121.124501} to nitrogen-vacancy center \cite{doi:10.1126/science.aaw8205} in solids and cold atoms \cite{Li2019,PhysRevLett.124.070402} and etc, which has facilitated investigations of novel physical phenomena including the real spectra in non-Hermitian Hamiltonians with parity-time ($\mathcal{PT}$) symmetry \cite{PhysRevLett.80.5243}, and non-Hermitian skin effect \cite{PhysRevLett.121.086803,Zhang31122022}, nontrivial non-Hermitian topology \cite{PhysRevLett.120.146402,Ghatak_2019,PhysRevX.9.041015,RevModPhys.93.015005,Ashida02072020,Okuma:2022bnb}. In the field of QPTs, non-Hermiticity gives rise to a new kind of phase transition, which is called the non-Hermitian QPTs and is characterized by the energy spectrum. This phase transition is closely related to $\mathcal{PT}$ symmetry and intrinsic rotation-time-reversal ($\mathcal{RT}$) symmetry \cite{PhysRevA.87.012114,PhysRevA.90.012103,PhysRevX.4.041001}. When one of the symmetries is conserved, the system has a pure real energy spectrum, whereas it becomes a complex energy spectrum when the symmetry is broken. The transition point is known as the exceptional point \cite{Heiss_2012}. At exceptional points, the non-Hermitian Hamiltonian is defective, and the eigenvectors will coalesce into one, where significant sensitivity enhancement at the exceptional points has been observed \cite{science_EPs}. 

The influence of non-Hermiticity on traditional QPTs has also attracted much interest, especially in spin models \cite{PhysRevA.105.053311,PhysRevB.104.155141,PhysRevB.110.014403,PhysRevA.110.012226,PhysRevLett.113.250401,PhysRevA.110.052203}. Some works studied QPTs under the influence of non-Hermiticity from the perspective of quantum information concepts such as entanglement \cite{10.21468/SciPostPhys.7.5.069,PhysRevResearch.2.033069,Guo:2021nmx, Chen_2024,PhysRevA.109.042208,PhysRevB.107.L020403,10.21468/SciPostPhys.14.5.138,soares2025symmetriesconservationlawsentanglement}, correlations \cite{PhysRevB.105.205125}, fidelity \cite{PhysRevA.98.052116,PhysRevA.105.053311,PhysRevResearch.3.013015, PhysRevA.110.052203,Tu:2022lvh}, coherence \cite{PhysRevLett.113.250401,PhysRevB.104.155141}, etc. In non-Hermitian systems, there exist two sets of eigenstates (the left and the right eigenstates) \cite{Brody_2014} which can be used to define two types of quantities, i.e., self-normal and biorthogonal quantities, to study QPTs \cite{10.21468/SciPostPhys.7.5.069}. It remains an open question whether the self-normal or the biorthogonal quantities are better indicators of the QPTs in non-Hermitian systems \cite{meden2023mathcal}. Recent research has shown that the biorthogonal fidelity susceptibility gives a more accurate Ising transition boundary than the self-normal one in the non-Hermitian transverse-field Ising model \cite{Sun2021}. 
%inadequacies in studying traditional QPTs under the influence of non-Hermiticiy in some models using self-normal quantities \cite{Sun2021}. 
Meanwhile, there are also some works studying the general properties of the biorthogonal quantities \cite{PhysRevA.98.052116,10.21468/SciPostPhys.7.5.069,Sun2021,PhysRevB.110.014441,Tu:2022lvh}. In particular, Ref. \cite{Tu:2022lvh} investigated the biorthogonal fidelity and its susceptibility and utilized them to study the $\mathcal{PT}$ and the first-order transitions in the XXZ model with the staggered imaginary field. However, these quantities fail to unveil the BKT transition of the model because of its slow divergence behavior with the system size. Therefore, the complete phase diagram of the model is still unclear. 

The study of non-Hermitian quantum phase transitions using both self-normal and biorthogonal quantum information tools deserves further investigation. In this paper, we investigate the QPTs of non-Hermitian spin models featuring exceptional points, with particular focus on the one-dimensional (1D) XXZ model and the transverse-field XY model from the perspective of quantum entanglement and correlations. We employ both self-normal and biorthogonal defined entanglement measures, including concurrence, quantum coherence, mutual information, and negativity, to characterize the phase transitions and examine how non-Hermiticity affects these transitions. In the XXZ model, we find that the critical points associated with the BKT transition and the first-order transition evolve into exceptional points as the parameter controlling non-Hermiticity increases. However, the first-order transition is more stable than the BKT transition in the $\mathcal{PT}$-symmetric regime in the sense that it persists in a stronger non-Hermiticity strength. A similar phenomenon occurs in the XY model, where the Ising transitions also transform into exceptional points with increasing non-Hermiticity. We find that all the quantum information measures considered, except for the biorthogonal concurrence, yield consistent phase boundaries of $\mathcal{PT}$ ($\mathcal{RT}$) transition and the first-order transition. To resolve this discrepancy, we propose the unconstrained concurrence and use it to successfully recover phase diagrams consistent with those obtained from other measures. Furthermore, for the BKT transition in the XXZ model, we find that the concurrence and negativity can give a consistent result with spin-spin correlation and fidelity-related methods at small non-Hermiticity strength, and other measures show no signal for the system size that we considered.

The rest of this paper is organized as follows. In Sec. \ref{sec2}, we introduce the methodology. In Sec. \ref{sec4}, we study the phase diagram of the non-Hermitian XXZ model by numerical calculations. In Sec. \ref{sec3}, we investigate the phase diagram of the non-Hermitian XY model by analytical calculations. Finally, we summarize the findings in Sec. \ref{sec5}.

\section{Methodology}\label{sec2}
In this section, we introduce several quantum information measures studied in this work. Let us begin with the two-site reduced density matrix $\rho_{i,j}$, which is obtained by tracing out all other degrees of freedom except that for the spins at sites $i$ and $j$. The two-site reduced density matrix takes the general form of \cite{Gu_2008}
\begin{eqnarray}
\rho_{i,j}&=&\frac{1}{4}+\frac{1}{4} \sum_\mu\left(\left\langle\sigma_i^\mu\right\rangle \sigma_i^\mu+\left\langle\sigma_j^\mu\right\rangle \sigma_j^\mu\right)\nonumber \\
&&+\frac{1}{4} \sum_{\mu v}\left\langle\sigma_i^\mu \sigma_j^\nu\right\rangle \sigma_i^\mu \sigma_j^\nu,
\end{eqnarray}
where $\sigma_i^{\mu}$ is the Pauli matrix with $\mu,\nu=x,y,z$. For systems possessing $Z_2$ symmetry, such as the XY model considered in the Sec. \ref{sec3}, the two-site reduced density matrix is simplified to
\begin{equation}
    \rho_{i,j}=\frac{1}{4}\left(\begin{array}{cccc}
     a_{11} & 0 & 0 & a_{14}\\
     0 & a_{22} & a_{23} & 0\\
     0 & a_{32} & a_{33} & 0\\
     a_{41} & 0 & 0 & a_{44}
\end{array}\right),
\label{eq:RDM2}
\end{equation}
where 
\begin{subequations}
\begin{align}
    a_{11} &= \braket{\sigma_i^z} + \braket{\sigma_j^z} + \braket{\sigma_i^z\sigma_j^z} + 1\\
    a_{22} &= \braket{\sigma_i^z} - \braket{\sigma_j^z} - \braket{\sigma_i^z\sigma_j^z} + 1\\
    a_{33} &= -\braket{\sigma_i^z} + \braket{\sigma_j^z} - \braket{\sigma_i^z\sigma_j^z} + 1\\
    a_{44} &= -\braket{\sigma_i^z} - \braket{\sigma_j^z} + \braket{\sigma_i^z\sigma_j^z} + 1\\
    a_{23} &= \braket{\sigma_i^x\sigma_j^x} +\braket{\sigma_i^y\sigma_j^y}+i\braket{\sigma_i^x\sigma_j^y}-i\braket{\sigma_i^y\sigma_j^x}\\
    a_{32} &= \braket{\sigma_i^x\sigma_j^x} +\braket{\sigma_i^y\sigma_j^y}-i\braket{\sigma_i^x\sigma_j^y}+i\braket{\sigma_i^y\sigma_j^x}\\
    a_{14} &= \braket{\sigma_i^x\sigma_j^x} - i\braket{\sigma_i^x\sigma_j^y} - i\braket{\sigma_i^y\sigma_j^x} - \braket{\sigma_i^y\sigma_j^y}\\
    a_{41} &= \braket{\sigma_i^x\sigma_j^x} + i\braket{\sigma_i^x\sigma_j^y}+i\braket{\sigma_i^y\sigma_j^x} - \braket{\sigma_i^y\sigma_j^y}.
\end{align}
\label{eq:RDM2_corr}
\end{subequations}
Based on the above two-site reduced density matrix, we can then calculate various measures of quantum entanglement and correlations, including mutual information, concurrence, negativity, and quantum coherence. Over the past few decades, these entanglement measures have been successfully applied to study QPTs in many Hermitian many-body systems without requiring prior knowledge of the order parameters \cite{YCLi2022_book}.

The mutual information measures the total correlation between two subsystems $i$ and $j$, and gives an upper bound of entanglement correlations \cite{PhysRevLett.100.070502,Gu_2008}. It is defined by 
\begin{equation}
    \mathcal{I}(\rho_{i,j}) = S(\rho_{i}) + S(\rho_j) - S(\rho_{i,j}),
    \label{eq:mutual_information}
\end{equation}
where $S(\rho_i)=-\text{Tr}(\rho_i\ln\rho_i)$ is the von-Neumann entropy of the subsystem $i$. Note that the entanglement entropy quantifies the resources needed to store a subsystem’s information. Equation (\ref{eq:mutual_information}) can be interpreted as the additional physical resources required to store the information of two subsystems separately instead of storing them together. In other words, it measures the correlation between the two subsystems. It has been shown that a non-vanishing mutual information between two subsystems separated far apart implies a non-vanishing long-range correlation \cite{PhysRevLett.100.070502,Gu_2008}. Based on the mutual information, a non-variational scheme to derive the potential order parameters by analyzing the reduced density matrix spectrum has been proposed \cite{Gu2013_OP}, and applied to several many-body systems \cite{Yu2016_hubbard,Yu2016_SSH,Yu2019,Yu2021}, including an interacting topological insulator \cite{Yu2016_SSH,Yu2021}.

The concurrence, on the other hand, can only be nontrivial if the system is entangled. It can be calculated by \cite{PhysRevLett.80.2245,PhysRevLett.78.5022}
\begin{equation}\label{eq4}
    C=\text{max}[0,\sqrt{\lambda_1}-\sqrt{\lambda_2}-\sqrt{\lambda_3}-\sqrt{\lambda_4}],
\end{equation}
where $\lambda_i (i=1,2,3,4)$ are the eigenvalues of a non-Hermitian matrix $R=\rho_{i,j}\tilde{\rho}_{i,j}$ and are sorted in the descending order. In the matrix $R$, $\tilde{\rho}_{i,j}=(\sigma^y_{i}\otimes\sigma^y_{j})\rho^{*}_{i,j}(\sigma^y_{i}\otimes\sigma^y_{j})$ is the spin-flipped density matrix, and $\rho^{*}_{i,j}$ is the complex conjugate of $\rho_{i,j}$. The concurrence was first applied to study the 1D transverse-field XY model in the context of QPTs, where its first derivative is found to be divergent at the quantum critical point \cite{Osterloh2002}. Subsequent studies on the XXZ model also found the concurrence to be a detector of the BKT transition in the model. However, instead of the singularity behavior observed in the second-order phase transition in the XY model, the concurrence attains a maximum at the BKT transition of the XXZ model \cite{Gu2003-jb,Gu2005-ak}. Recently, the concurrence has also been examined in the non-Hermitian XY model with $\mathcal{RT}$-symmetry, where it is found to be maximum at the exceptional points \cite{PhysRevB.110.014403}.

We also consider logarithmic negativity\cite{PhysRevLett.77.1413,HORODECKI19961,PhysRevA.65.032314,PhysRevLett.95.090503}, which is defined as 
\begin{equation}
    \varepsilon(\rho_{i,j})=\log_2(2\mathcal{N}(\rho_{i,j})+1),
\end{equation}
where $\mathcal{N}$ is the absolute sum of negative eigenvalues of the partially transposed reduced density matrix. Negativity has become a widely used entanglement measure in mixed states because it is sensitive only to genuine quantum correlations but unaffected by thermal fluctuations. Researchers have applied it to various Hermitian systems, including fermionic systems \cite{Eisler_2015,PhysRevB.95.165101}, spin systems \cite{PhysRevB.94.035152,Mbeng_2017}, and one-dimensional conformal field theories \cite{Calabrese_2015}. Moreover, negativity has been shown to distinguish finite-temperature phase transitions, particularly in phases with spontaneous symmetry breaking \cite{PhysRevB.99.075157}, and in systems exhibiting topological order \cite{PhysRevLett.125.116801}. It has also been applied to identify phase transitions in the non-Hermitian XY spin chain with Hermitian Kaplan–Shekhtman–Entin-Aharony interactions under a transverse magnetic field \cite{Agarwal:2023nvj}.

For quantum coherence, we study the quantum coherence based on the Wigner-Yanase skew information which quantifies the amount of information contained in a quantum state and reflects the information of a state skewed to an observable \cite{Wigner:1963zuk}. It has been proven to satisfy all the criteria for coherence monotones \cite{PhysRevLett.113.140401} and can be used as an efficient measure to quantify quantum coherence. The skew information has been studied to reveal the occurrence of QPTs \cite{PhysRevB.90.104431, Li2016}. Here we consider the lower bound of Wigner-Yanase skew information \cite{PhysRevLett.113.170401}, it is defined as follows:
\begin{equation}
    QC(\rho_{i,j},O) = -\frac{1}{4}\text{Tr}([\rho_{i,j},O]^2),
\end{equation}
where $O=O_{i}\otimes I_{j}$ is an observable if we choose the observable at site $i$, and $[.,.]$ denotes the commutator. In the following, we use the observable $O=\sigma^x_i\otimes I_j$ and we investigate the quantum information measures between two neighboring sites by choosing $i=L/2$ and $j=L/2+1$. A recent study showed that the quantum coherence, particularly the second moment of multiple quantum coherence intensities, can signal critical points in several paradigmatic non-Hermitian Hamiltonians \cite{PhysRevB.104.155141}.

As the Hamiltonian is non-Hermitian, the density matrix can be defined in different ways, including left and right eigenvectors. Here, we adopt two definitions of the density matrix, i.e. 
$\rho^{RR}=\frac{\ket{G_R}\bra{G_R}}{\braket{G_R|G_R}}$ and  $\rho^{RL}=\frac{\ket{G_R}\bra{G_L}}{\braket{G_L|G_R}}$, where $\ket{G_R}$($\ket{G_L}$) is the right (left) ground state. The reduced density matrix in Eq. (\ref{eq:RDM2}) and the entanglement quantities will then have the corresponding self-normal and biorthogonal definitions.

\section{XXZ model}\label{sec4}
\begin{figure}[!tb]
    \centering\includegraphics[width=.45\textwidth]{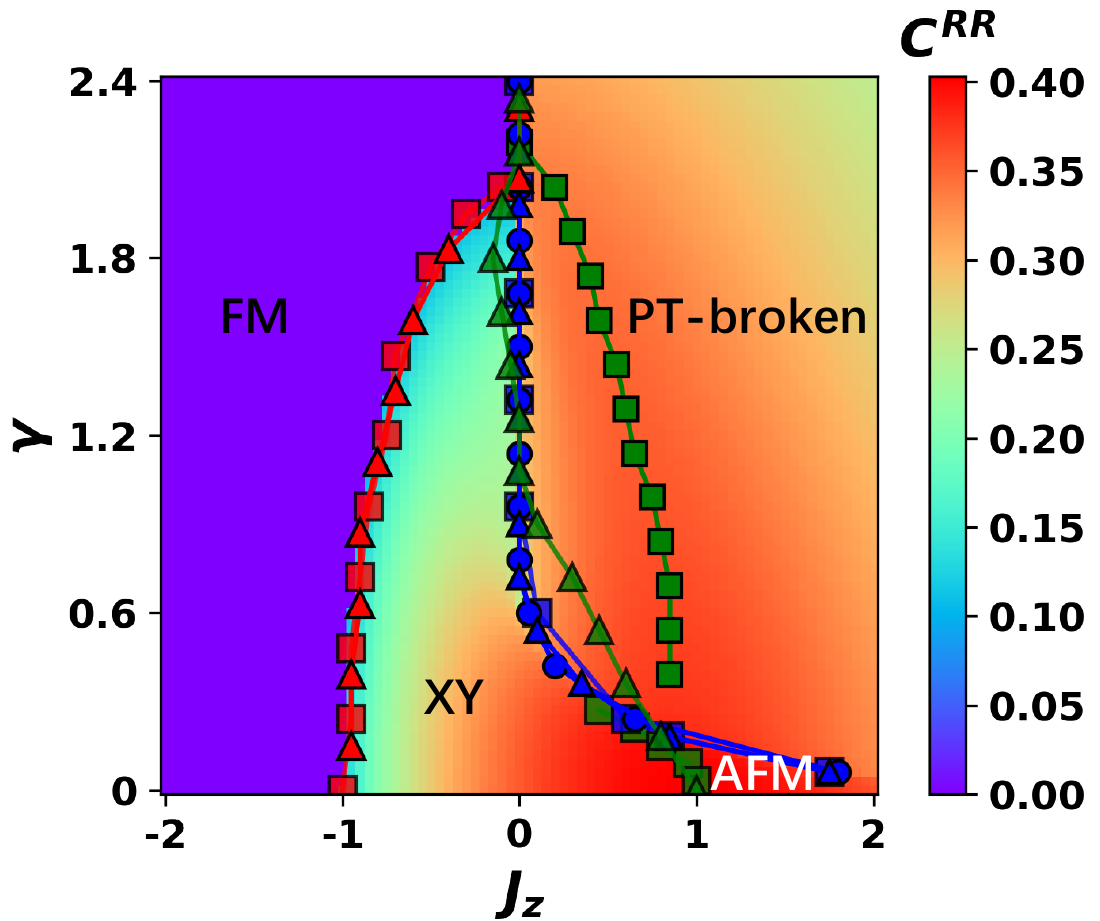}
    \caption{Phase diagram of the non-Hermitian XXZ model on the $\gamma-J_z$ plane for system size $N=10$. The color scale shows the magnitude of the concurrence. The first-order transition line between the FM and XY phases is determined from the discontinuous jump in the concurrence (red squares) and in the spin-spin correlation functions (red triangles), which agree well with each other. The BKT transition between the XY and the AFM phases is determined by the maximum of the concurrence (green squares) and the crossing of the spin-spin correlation functions (green triangles), which agree with each other in small $\gamma$. The $\mathcal{PT}$ transition is determined by the discontinuity in the concurrence (blue squares) and the spin-spin correlation (blue triangles), and directly from the energy spectrum (blue circles).}
    \label{fig1}
\end{figure} 
\begin{figure*}[!tb]
    \centering\includegraphics[width=.88\textwidth]{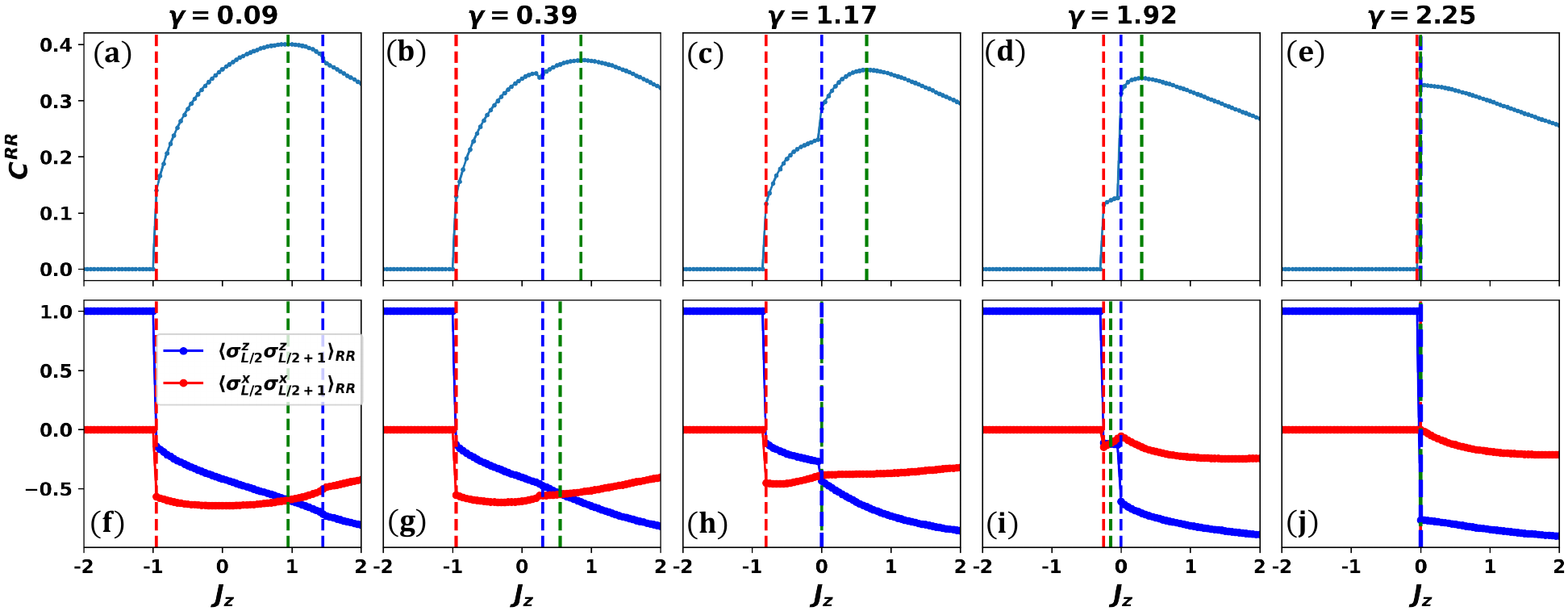}
    \caption{Concurrence (a-e) and spin-spin correlation (f-j) as a function of $J_z$ for five typical $\gamma$ values. (a-e) The red and blue dashed lines indicate the first-order and the $\mathcal{PT}$ transitions as obtained by the extrema of the concurrence's first derivative. The green dashed lines indicate the BKT transition obtained from the maximum of the concurrence. (f-g) The red and blue dashed lines are obtained by the extrema of the first derivative of $\braket{\sigma_{L/2}^z\sigma_{L/2+1}^z}_{RR}$, and the green dashed lines are obtained by the crossing points of spin-spin correlations.}
    \label{fig2}
\end{figure*} 
Let us begin with the non-Hermitian spin-$\frac{1}{2}$ XXZ model defined by the Hamiltonian
\begin{equation}
    H=\sum_{l=1}^N (\sigma_l^x\sigma_{l+1}^x+\sigma_l^y\sigma_{l+1}^y+J_z\sigma_l^z\sigma_{l+1}^z)+i\gamma\sum_{l=1}^{N/2}(\sigma_{2l-1}^z-\sigma_{2l}^z),
\end{equation}
where $\gamma$ controls the strength of the staggered imaginary magnetic field along the $z$-direction. In the Hermitian limit ($\gamma=0$), the ground state phase diagram is determined by the anisotropy term $J_z$. When $J_z\rightarrow-\infty$ ($\infty$), the spin-spin interaction along the $z$ direction is dominant, and the system is in the ferromagnetic (antiferromagnetic) phase. Between these two phases, the first two exchange terms introduce quantum fluctuations, resulting in the XY (also called the Luttinger liquid) phase. At $J_z=-1$, the system undergoes a first-order transition between the ferromagnetic (FM) and the XY phases, while at $J_z=1$, the system undergoes a BKT transition between the XY and the antiferromagnetic (AFM) phases. When the non-Hermitian term is turned on ($\gamma\neq0$), the model has $\mathcal{PT}$ symmetry with $\mathcal{T}i\mathcal{T} = -i$ and $\mathcal{P}\sigma^{\alpha}_{j}\mathcal{P}=\sigma^{\alpha}_{N+1-j}$. The $\mathcal{PT}$ transition has been identified by the divergence of the real part of biorthogonal fidelity susceptibility \cite{Tu:2022lvh}.
\begin{figure*}[!tb]
    \centering\includegraphics[width=.88\textwidth]{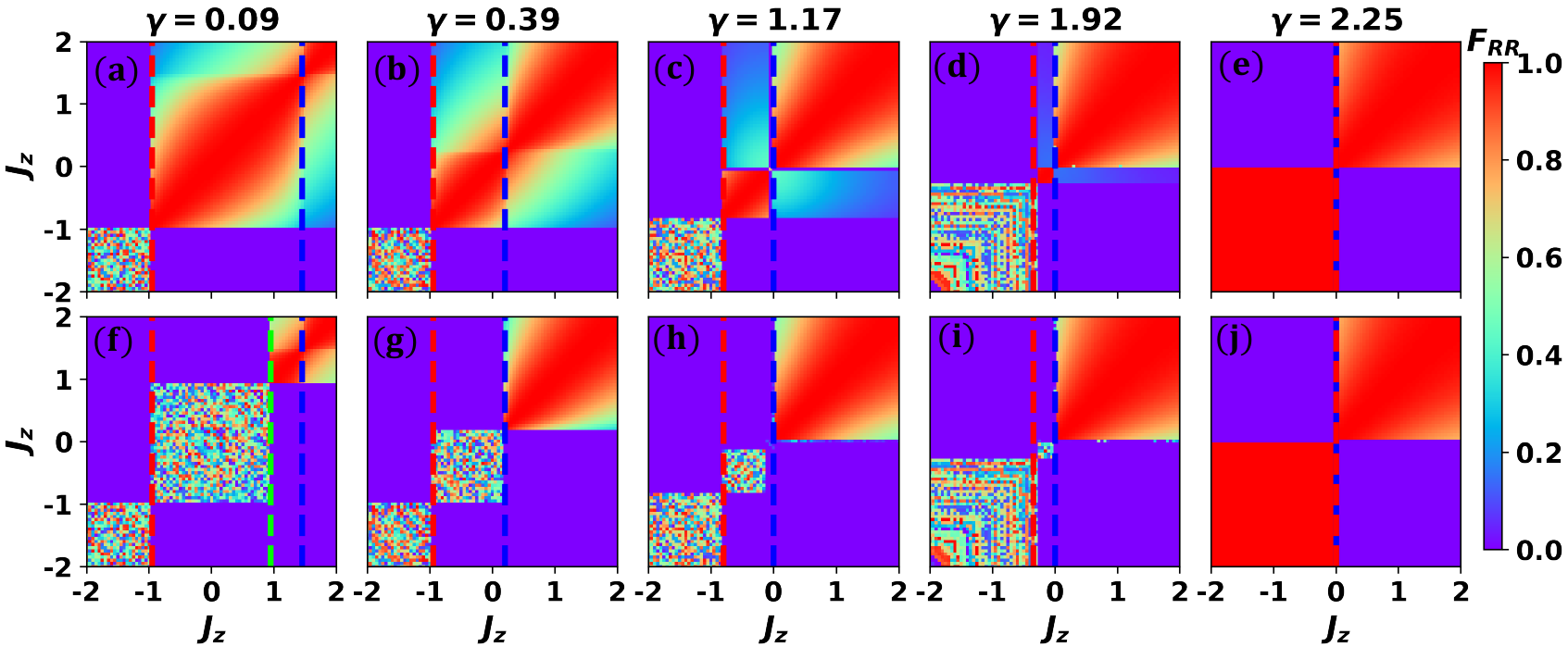}
    \caption{The ground state (a-e) and first excited state (f-j) fidelity map of five typical values of $\gamma$. The red, green, and blue dashed line denotes the first-order, BKT, and $\mathcal{PT}$ transitions, respectively, as determined from the self-normal concurrence.}
    \label{fig3}
\end{figure*} 
\begin{figure*}[!tb]
    \centering\includegraphics[width=.88\textwidth]{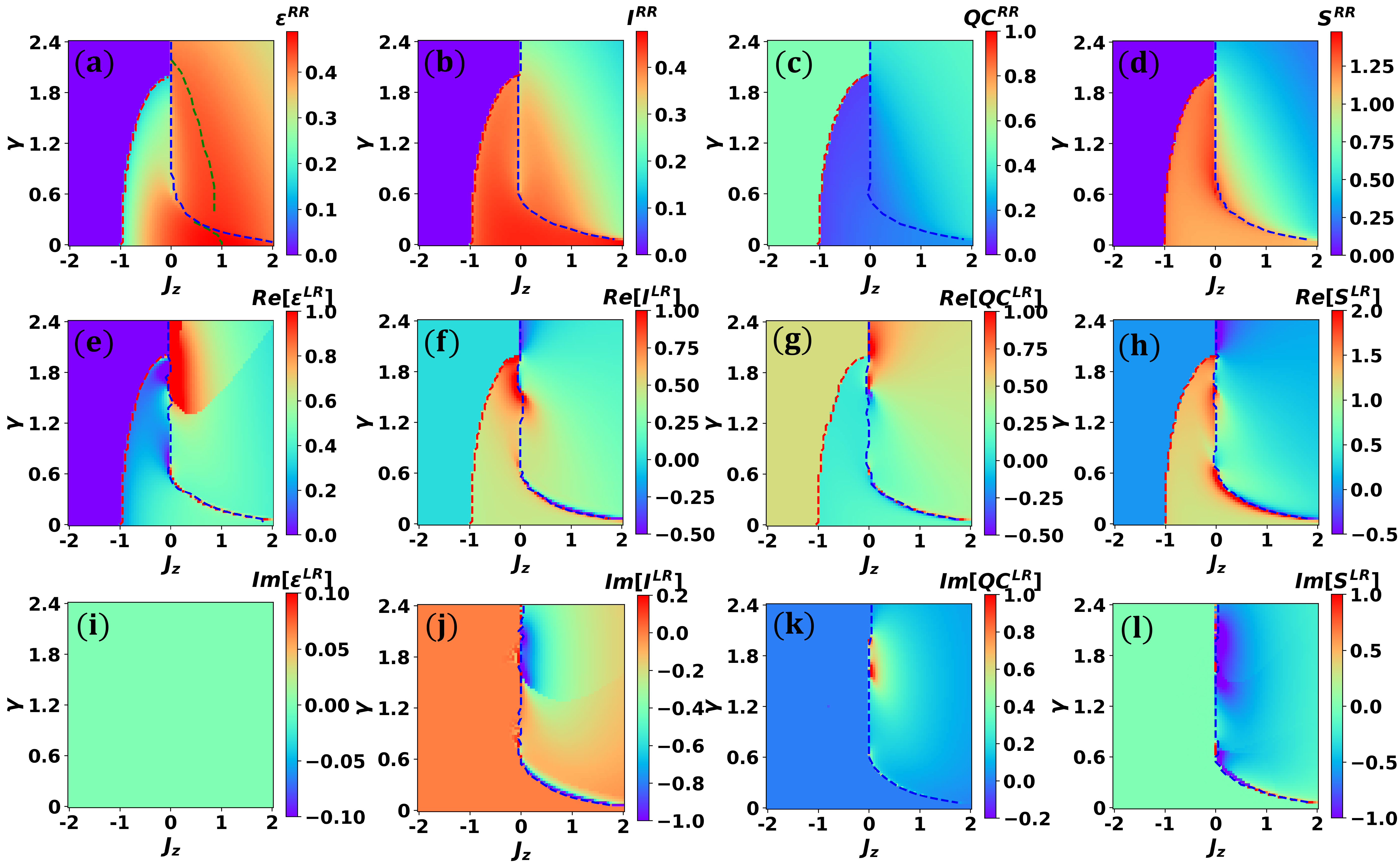}
    \caption{The negativity (first column), mutual information (second column), quantum coherence (third column), and half-chain entanglement entropy (fourth column) as a function of the imaginary field strength $\gamma$ and the anisotropy strength $J_z$ of the XXZ model. The top, middle, and bottom panel shows the self-normal, the real part of the biorthogonal, and the imaginary part of the biorthogonal quantities, respectively. The dashed lines indicate the phase boundaries determined by the extrema or the discontinuity of the concerned quantities.}
    \label{fig4}
\end{figure*} 
\begin{figure*}[!tb]
    \centering\includegraphics[width=.88\textwidth]{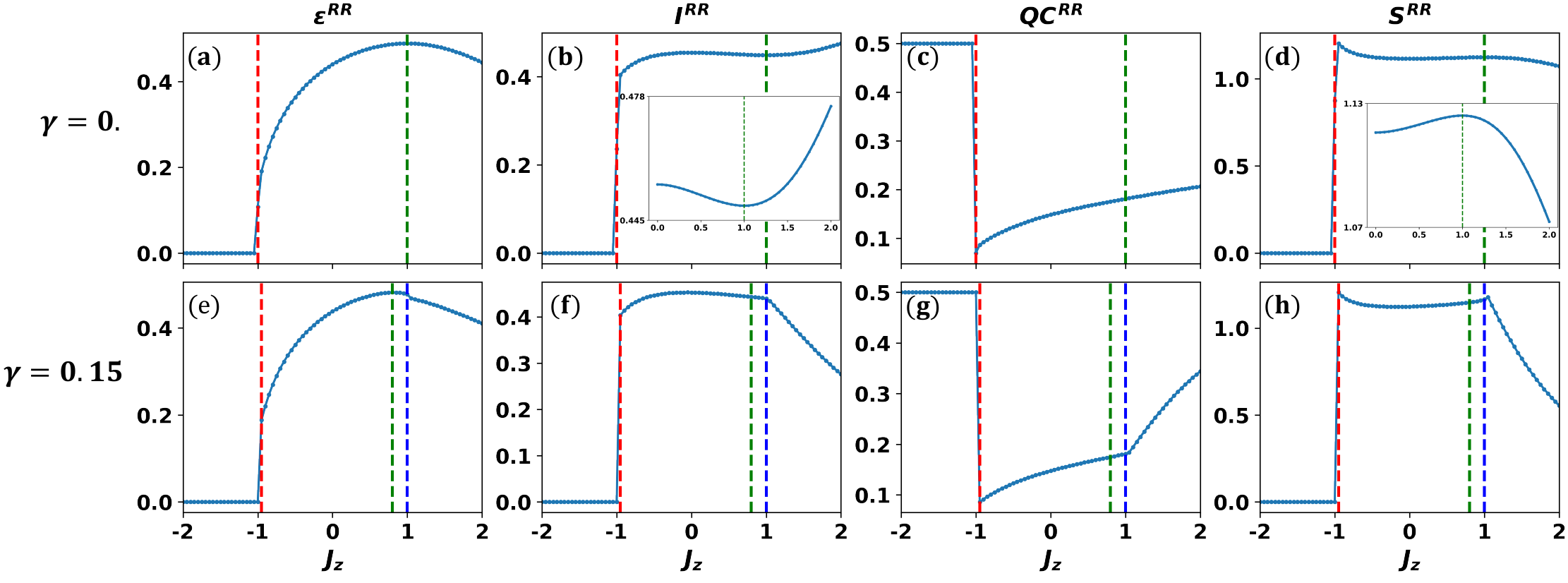}
    \caption{The self-normal entanglement measures as a function of $J_z$ for $\gamma=0$ (top panel) and $\gamma=0.15$ (bottom panel). The red, green and blue dashed lines indicate the first-order, BKT, and $\mathcal{PT}$ transition, respectively, as determined from the self-norm concurrence. All four measures capture the first-order and $\mathcal{PT}$ transitions of the system, but only the negativity detects the BKT transition by its maximum. Insets show a zoom-in of the plot in the vicinity of the BKT transition around $J_z=1$.}
    \label{fig5}
\end{figure*} 
\begin{figure}[!tb]
    \centering\includegraphics[width=.45\textwidth]{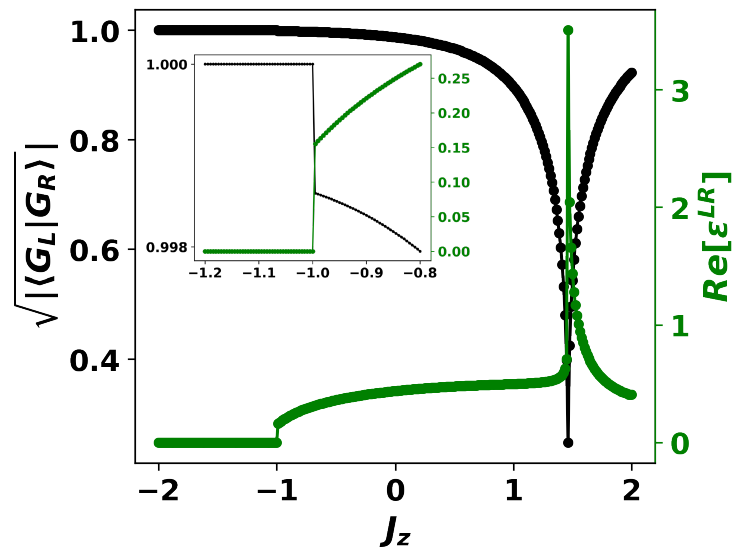}
    \caption{The normalization factor of the ground state (left y-axis) and the real part of the biorthogonal negativity (right y-axis) as a function of $J_z$ for $\gamma=0.09$ in the XXZ model. The normalization factor exhibits a singularity, which in turn reflects in the negativity around the exceptional point. The inset shows a zoom-in of the plot in the vicinity of the first-order transition around $J_z = -1$.}
    \label{fig6}
\end{figure} 
\begin{figure*}[!tb]
    \centering\includegraphics[width=.88\textwidth]{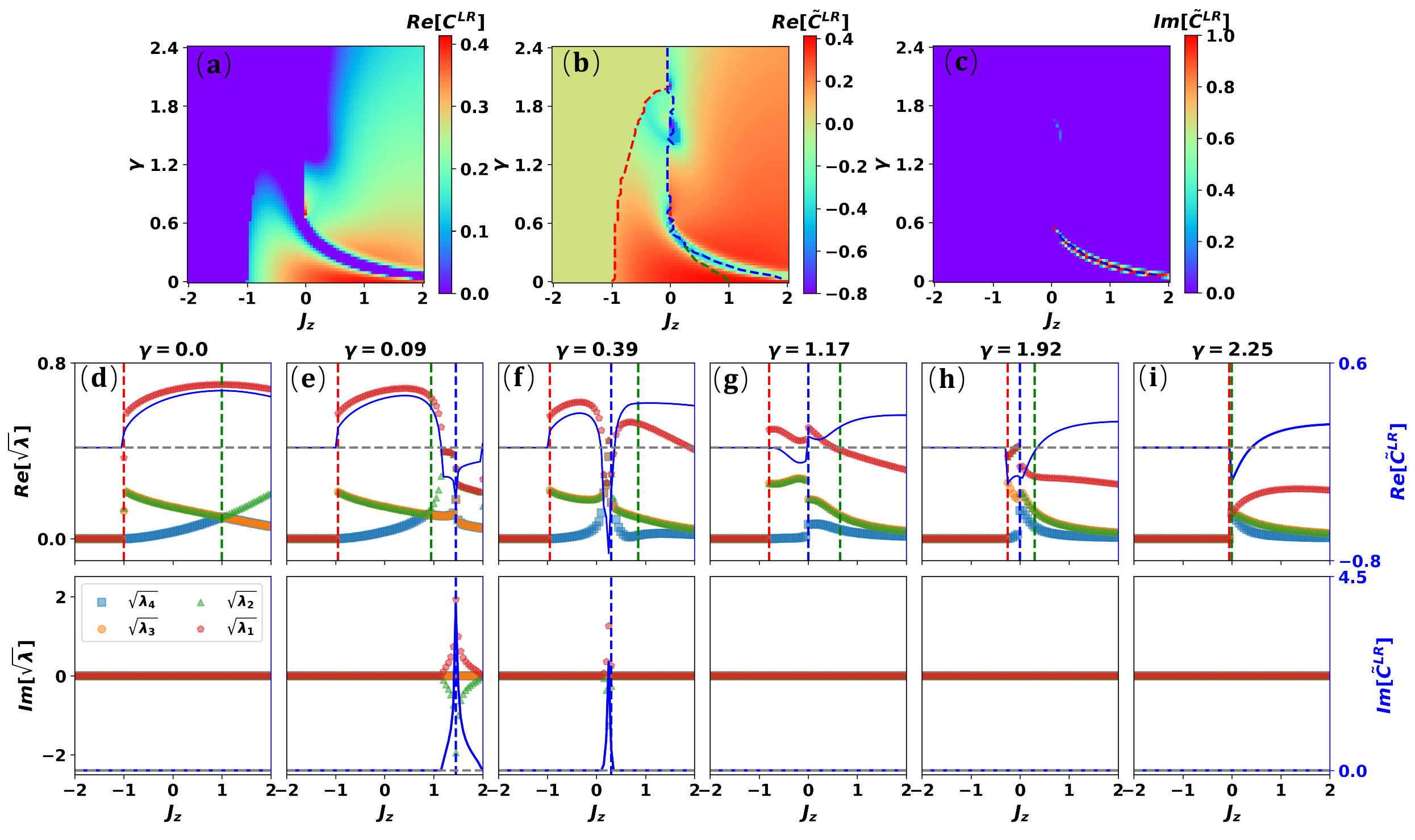}
    \caption{(a) The real part of the biorthogonal concurrence, and (b-c) the biorthogonal unconstrained concurrence as a function of $\gamma$ and $J_z$. The red and blue dashed line are determined by the discontinuity and extrema of $Re[\tilde{C}^{LR}]$, respectively, while the green dashed line is determined by the degeneracy of the second and third largest eigenvalues. (d-i) The eigenvalue spectrum of the $R$ matrix defining the concurrence in Eq. \ref{eq4} (left y-axis) and the biorthogonal unconstrained concurrence (right y-axis) as a function of $J_z$ for six typical $\gamma$ values. The horizontal dashed lines indicate the zero of the right y-axis. The red, green, and blue dashed lines in (d-i) are the first-order, BKT, and $\mathcal{PT}$ transition points determined from self-normal concurrence, respectively.}
    \label{fig7}
\end{figure*} 

In the following, we consider the eigenstate that has the minimum real part of the eigenenergy as the ground state \cite{Liu_2021} and a system of $N=10$ spins unless otherwise specified. The ground state phase diagram on the $\gamma-J_z$ plane is shown in Fig. \ref{fig1}. We use four quantities, including self-normal concurrence $C^{RR}$, spin-spin correlation, the ground state energy and the fidelity map \cite{PhysRevB.104.075142}, to characterize the QPTs. In the Hermitian case, as shown in previous studies \cite{Gu2003-jb}, the discontinuity and the maxima of the concurrence well capture the first-order ($J_z=-1$) and BKT transitions ($J_z=1$) in the model, respectively. It has also been found that the spin-spin correlation function along the $x$ (or $y$) direction and the $z$ direction crosses at the BKT transition where the spin interactions become isotropic \cite{Werlang2010-vx}. As the non-Hermiticity turns on, for small $\gamma$, we find that the concurrence shows similar features and it detects the first-order and BKT transitions by its discontinuity jump and maximum, respectively, as shown in Fig. \ref{fig2} (a). In addition, Fig. \ref{fig2} (f) shows the self-normal spin-spin correlations along the $x$ and $z$ directions. The correlation functions show a discontinuous jump around $J_z=-1$ and a crossing around $J_z=1$, aligning with the first-order and BKT transition points detected by the concurrence. While the alignment in the transition, detected by the concurrence and the spin correlations, holds for the first-order transition across a considerable range of $\gamma\lessapprox 2$, the alignment in detecting the BKT transition only holds for small $\gamma\lessapprox 0.28$. For intermediate values of $\gamma$, the maximum of concurrence does not agree with where the spin correlations cross (Figs. \ref{fig2} (b-e) and (g-j)). Moreover, as the non-Hermiticity turns on, the system also exhibits a $\mathcal{PT}$ transition where the ground state energy changes from completely real to complex. As shown by the blue dashed lines in Fig. \ref{fig2}, this $\mathcal{PT}$ transition can be identified by the cusps in the concurrence and the spin-spin correlations.

We further benchmark the results from the concurrence and spin-spin correlations using the fidelity map proposed in Ref. \cite{PhysRevB.104.075142}. The fidelity map approach utilizes the overlap between different ground (or excited) states within a parameter range to construct a two-dimensional map, which captures more comprehensive information about the system compared to conventional fidelity methods that rely solely on neighboring states, to study QPTs. Using the similarity of states,  if the ground state across the QPT is completely different, the fidelity map will show a sharply distinguishable region. However, if the two phases concerned share some similar features, the transition would be smooth on the fidelity map. Figures \ref{fig3} (a-e) show the ground state fidelity map which captures both the first-order and the $\mathcal{PT}$ transitions that are consistent with the transition points signaled by concurrence and the spin-spin correlations. As the non-Hermiticity $\gamma$ increases, the first-order transition (red dashed line) and $\mathcal{PT}$ transition (blue dashed line) approach $J_z=0$ and eventually merge to a single transition at large $\gamma$. Figure \ref{fig3}(f-j) shows the fidelity map of the first excited state. It has been demonstrated in the Hermitian models that the first excited state fidelity is a more effective indicator of the BKT transition attributed to the level crossing between the first and second excited states \cite{PhysRevB.104.075142,Chen2007-fh}. In addition to the first-order and $\mathcal{PT}$ transitions in the non-Hermitian XXZ model, the first excited state fidelity map further captures the BKT transition (green dashed line) in small $\gamma$. The noisy feature in the fidelity map is a result of the degeneracy in the ground state or the first excited state.

The general result of the phase diagram is that when $\gamma$ increases, the first-order transition and the BKT transition merge into the $\mathcal{PT}$ transition. To be specific, at small $\gamma$ ($\gamma\lessapprox 0.28$), we find that the results of the four studied quantities are consistent. They all indicate the occurrence of a $\mathcal{PT}$ transition and a BKT transition in the $\mathcal{PT}$-symmetric regime stemming from $J_z = 1$ in the Hermitian case. The two transitions approach each other as non-Hermiticity increases and meet at $\gamma\approx 0.28$. This result suggests that the XY phase shrinks as non-Hermiticity increases. This behavior is opposite to recent results on the XXZ model with complex Dzyaloshinskii-Moriya interaction, which can be mapped to a bosonic model with nearest-neighbor non-reciprocal hopping~\cite{PhysRevA.109.042208}. On the other hand, the first-order and the $\mathcal{PT}$ transitions approach each other as $\gamma$ increases and eventually merge into a single one for $\gamma \gtrapprox2$. At intermediate values of $\gamma$ ($0.28\lessapprox \gamma  \lessapprox 2$), the maximum of $C^{RR}$ and the crossing of the spin-spin correlation give an inconsistent prediction of the potential BKT transition point, and there is no signal of the transition observed in the fidelity map. It is unclear whether the BKT transition is present in this case, but if it exists, it will likely be in the $\mathcal{PT}$-broken regime and the signal of spin-spin correlation and $C^{RR}$ may have been affected by the singularity of the $\mathcal{PT}$ transition. At large $\gamma$($\gamma\gtrapprox 2$), there only exists the $\mathcal{PT}$ transition. 

Figure \ref{fig4} further shows other self-normal and bi-orthogonal quantum information measures, including negativity, mutual information, quantum coherence, and half-chain entanglement entropy, as discussed in Sec. \ref{sec2}. We find that the self-normal and the real part of the biorthogonal quantities can signal the first-order and $\mathcal{PT}$ transitions, and the results are consistent with the phase diagram shown in Fig. \ref{fig1}. On the other hand, the imaginary part of the biorthogonal quantities except for the negativity, which is always real due to the absolute sum in its definition, can only detect the $\mathcal{PT}$ transition. Furthermore, among the quantities concerned, only the self-normal negativity signals the BKT transition at small $\gamma$. This is evidence from Fig. \ref{fig5} where the self-normal measures as a function of $J_z$ at $\gamma=0$ (the Hermitian case) and $\gamma=0.15$ are displayed. The maximum of the negativity agrees with the BKT transition points determined from the self-normal concurrence. For mutual information and half-chain entanglement entropy, the insets show that the signal of the BKT transition is weak due to the finite size effect. To capture the signal of the BKT transition for other quantities, one may need to resort to a large system, which is inaccessible for the exact diagonalization \cite{PhysRevA.81.064301}. Large-scale numerical simulations, such as the density matrix renormalization group and the quantum Monte Carlo method, are an active research direction for non-Hermitian models \cite{PhysRevLett.130.100401,PhysRevB.101.235150,PhysRevB.105.205125,PhysRevLett.132.116503,Guo:2022yfi,zhong2024DMRG}. Besides, we discuss the deficiency from the definition of the measures in capturing the BKT transition by investigating the eigenvalue spectrum of the respective reduced density matrix in Appendix \ref{appA}. Alternatively, the string order parameter \cite{PhysRevB.40.4709,PhysRevB.78.224402} may be a possible candidate of the order parameter. However, it is a non-local quantity, and concerning the small system size achievable in the ED simulation, we may not be able to obtain meaningful results. It would be an interesting future work with advancements in other simulation methods capable of tackling larger systems.

The sensitivity of the biorthogonal quantities in detecting the $\mathcal{PT}$ transition can be attributed to the singular drop in the normalization factor used to normalize the ground states, i.e. $ \bra{\tilde{G}_L}=\frac{\bra{G_L}}{\sqrt{\braket{G_L|G_R}}}$ and $\ket{\tilde{G}_R}=\frac{\ket{G_R}}{\sqrt{\braket{G_L|G_R}}}$, where $ \bra{\tilde{G}_L}$ and $\ket{\tilde{G}_R}$ are the normalized ground states that satisfy the biorthogonal condition, as shown in Fig. \ref{fig6}. Due to the self-orthogonality \cite{moiseyev2011non,Heiss_2012} at the exceptional point, as the system size increases, the normalization factor should approach zero. A similar result has also been reported in the non-Hermitian transverse-field Ising model \cite{PhysRevB.110.014441}. The inset shows that the normalization factor has a discontinuous drop in the vicinity of the first-order transition. The behavior around the transition points appears to be model-dependent. It would be interesting future work to investigate whether any universal behavior exists in how the normalization factor or other relevant quantities approach exceptional points. One may naturally expect that such an abrupt change can also be reflected in the biorthogonal concurrence. However, we find that this is not the case. The biorthogonal concurrence is featureless around the $\mathcal{PT}$ transition, especially in large $\gamma$, as shown in Fig. \ref{fig7} (a). The reason for this is the largest and the second largest eigenvalues, i.e. $\lambda_1$ and $\lambda_2$, in Eq. (\ref{eq4}) are degenerated around the exceptional point, and hence $\sqrt{\lambda_1}-\sqrt{\lambda_2}-\sqrt{\lambda_3}-\sqrt{\lambda_4}$ becomes negative (see Fig. \ref{fig7} (e)). Therefore, the concurrence defined in Eq. (\ref{eq4}) becomes zero and masks the signal of the $\mathcal{PT}$ transition. 

To unveil the signal of the $\mathcal{PT}$ transition, we propose the biorthogonal unconstrained concurrence 
\begin{eqnarray}
\tilde{C}^{LR} = \sqrt{\lambda_1}-\sqrt{\lambda_2}-\sqrt{\lambda_3}-\sqrt{\lambda_4}.
\label{eq:newC}
\end{eqnarray}
This redefined concurrence successfully captures the phase transitions as shown in Fig. \ref{fig7} (b-c): its real component signals both first-order and $\mathcal{PT}$ transitions (see  Fig. \ref{fig7} (d-i)), while its imaginary component partially detects the $\mathcal{PT}$ (see Fig. \ref{fig7} (d-i)). Besides, we find the degeneracy of the second and third largest eigenvalues reveals the BKT transition in small $\gamma$, which agrees with the result of self-normal concurrences, shown in Fig. \ref{fig7} (b), (d-e).

\section{non-Hermitian XY model}\label{sec3} 
We further consider the non-Hermitian XY model. Building on the previous analytical solution for self-normalized measures \cite{PhysRevX.4.041001,PhysRevB.110.014403}, we extend the framework to biorthogonal measures. We then investigate biorthogonal quantum information measures and examine their consistency with self-normalized measures in predicting the phase diagram, which remains an open question. The Hamiltonian reads as \cite{PhysRevA.87.012114} 
\begin{equation}
    H = -\sum_{l=1}^{N} (\frac{1+i\gamma}{2}\sigma_l^x\sigma_{l+1}^x + \frac{1-i\gamma}{2}\sigma_l^y\sigma_{l+1}^y + h\sigma_l^z),
    \label{eq:H_xy}
\end{equation}
where $\sigma_l^x$, $\sigma_l^y$ and $\sigma_l^z$ are Pauli matrices of the $l$-th spin. $\gamma$ is a real number and it measures the non-Hermitian anisotropy between $x$ and $y$ couplings, $i$ is the imaginary unit, $h$ is the transverse external magnetic field, lying along the $z$-direction and we impose periodic boundary conditions. This model has $\mathcal{RT}$ symmetry, where the linear rotation operator $\mathcal{R}\equiv \exp{[-i(\pi/4)\sum_{l=1}^N \sigma_l^z]}$ rotates each spin by $\pi/2$ about the $z$ axis, and the antilinear time-reversal operator $\mathcal{T}$ has the function $\mathcal{T}i\mathcal{T}=-i$.  The Hamiltonian features a pure real energy spectrum in the symmetry-preserving region, whereas it possesses a complex energy spectrum in the region of broken symmetry.

The Hamiltonian in Eq. (\ref{eq:H_xy}) can be diagonalized through standard procedures \cite{PhysRevA.87.012114,PhysRevX.4.041001,mbengQuantumIsingChain2024,PhysRevB.110.014403}. First, we rewrite the Hamiltonian in the spinless fermionic representation via Jordan-Wigner transformation, which is defined as 
\begin{subequations}
\begin{align}
    \sigma_l^z &= 2c_l^\dagger c_l-1,\\
    \sigma_{l}^{-} &= \prod_{j<l}(1-2c_j^\dagger c_j)c_l,\\
     \sigma_{l}^{+} &= \prod_{j<l}(1-2c_j^\dagger c_j)c_l^\dagger,
\end{align}
\end{subequations}
where $\sigma_l^{\pm}=\frac{1}{2}(\sigma_l^x\pm i\sigma_l^y)$, $c_l^\dagger$ and $c_l$ are the creation and annihilation operators at site $l$, respectively.  Then a Fourier transformation with $c_l=\frac{e^{-i\pi/4}}{\sqrt{N}}\sum_ke^{ikl}c_k$ is performed and the Hamiltonian in momentum space can be written as
\begin{align}
    H =& \sum_{k\in K}[(\cos k + h)c_{-k}c_{-k}^\dagger-(\cos k +h)c_k^\dagger c_k \notag\\ &+ i\gamma \sin k c_{-k}c_k + i\gamma \sin k c_{k}^\dagger c_{-k}^\dagger],
    \label{eq:Hk_xy}
\end{align}
where the even parity sector which applies the anti-periodic boundary conditions $k\in K=\{\pm\frac{(2n-1)\pi}{N},n=1,...,N/2\}$ is chosen. The Hamiltonian in Eq. (\ref{eq:Hk_xy}) can also be expressed as $H =\sum_{k\in K}\left(\begin{array}{cc}
     c_k^\dagger& c_{-k}
\end{array}\right)\mathcal{H}(k) \left(\begin{array}{c}
     c_k\\
     c_{-k}^{\dagger}
\end{array}\right)$ with $\mathcal{H}(k)=(i\gamma\sin k)\sigma_x-(\cos{k}+h)\sigma_z$ is the Bogoliubov-de Gennes (BdG) Hamiltonian. Then the Hamiltonian can be diagonalized through the Bogoliubov transformation into
\begin{equation}
    H =\sum_{k\in K,k>0}\epsilon_k(\bar{\eta}_k\eta_k+\bar{\eta}_{-k}\eta_{-k}-1),
\end{equation}
where $\epsilon_k=\sqrt{(\cos{k}+h)^2-\gamma^2\sin^2{k}}$ is the energy of each mode and it is symmetric about $k=0$, i.e. $\epsilon_k=\epsilon_{-k}$. Here, $\eta_k$ and $\bar{\eta}_{k}$ are non-Hermitian Bogoliubov quasiparticles, which can be defined as,
\begin{subequations}
\begin{align}
    \eta_k&=u_k c_k + v_k c_{-k}^\dagger, \quad \eta_{-k}=-v_k c_k^\dagger + u_k c_{-k},\\
    \bar{\eta}_k&=u_k c_k^\dagger + v_k c_{-k},\quad \bar{\eta}_{-k}=-v_k c_k + u_k c_{-k}^\dagger.
\end{align}   
\end{subequations}
We parametrize 
\begin{subequations}
\begin{align}
    u_k &= \frac{-\cos{k}-h\pm\sqrt{(\cos{k}+h)^2-\gamma^2\sin^2{k}}}{\mathcal{C}},\\
    v_k &= \frac{i\gamma \sin{k}}{\mathcal{C}}, 
\end{align}
\end{subequations}
where the normalization constant $\mathcal{C}$ is such that $u_k^2+v_k^2=1$. Since $u_k,v_k$ are complex numbers, $\eta_k^\dagger\neq\bar{\eta}_k$. The fermionic anti-commutation relations still hold, i.e., $\{\bar{\eta}_k,\eta_{k'}\}=\delta_{kk'}$ and $\{\eta_k,\eta_{k'}\}=\{\bar{\eta}_k,\bar{\eta}_{k'}\}=0$. 

The ground state of the model is given by
\begin{equation}
    \ket{G_R} = \frac{1}{\sqrt{\mathcal{N}}}\prod_{k>0}[u_k - v_k c_k^\dagger c_{-k}^\dagger]\ket{\text{Vac}},
\end{equation}
where $\ket{\text{Vac}}$ denotes the vacuum state of the free fermion and $\mathcal{N}=\prod_{k>0}(|u_k|^2+|v_k|^2)$ is  the normalization constant. The eigenvalue of $ \ket{G_R}$ is $E_0=-\sum_k \epsilon_k$,  and we choose the sign convention such that both the real and the imaginary parts of the ground state energy are the lowest since $\epsilon_k$ is either purely real or imaginary. The corresponding left ground state can be obtained by $(\bar{\eta}_k)^\dagger\ket{G_L}=0$, and we have $\ket{G_L}=\sqrt{\mathcal{N}}\prod_{k>0}(u_k^*-v_k^* c_k^\dagger c_{-k}^{\dagger})\ket{\text{Vac}}$.

\begin{figure*}[!htb]
    \centering
    \includegraphics[width=.88\textwidth]{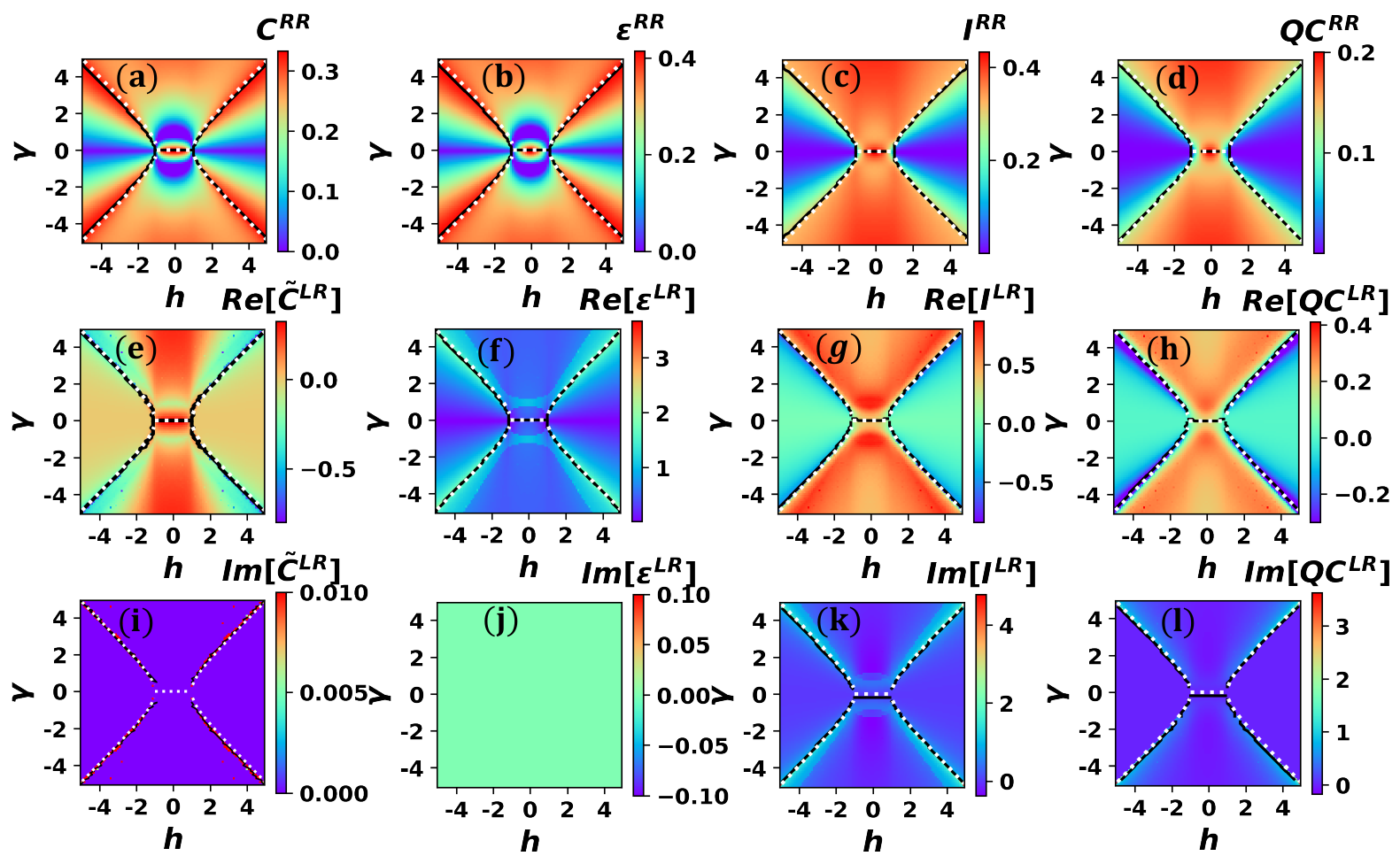}
    \caption{Phase diagram of the non-Hermitian XY model on $\gamma-h$ plane. The four quantum information quantities with self-normal and biorthogonal versions. Their derivatives can signal precisely the presence of a quantum phase transition at the critical points. The black lines are the phase boundary obtained by the data; white dotted lines are the analytical solution.}
    \label{fig8}
\end{figure*}    

To calculate the self-normal expectation values in Eq. (\ref{eq:RDM2_corr}), we consider $\eta_k$ and $\eta_k^\dagger$ instead of $\eta_k$ and $\bar{\eta}_k$. Note that $c_k=\frac{u_k^*\eta_k-v_k\eta_{-k}^\dagger}{|u_k|^2+|v_k|^2}$ and $\braket{\eta_k \eta_k^\dagger}=|u_k|^2+|v_k|^2
$, the non-vanishing correlation functions for the ground state take the following form in the momentum space:
\begin{subequations}
\begin{align}
    \braket{c_p^\dagger c_k^\dagger}&=\delta_{p,-k}\frac{-u_k v_p^*\braket{\eta_{-p} \eta_k^\dagger}}{(|u_p|^2+|v_p|^2)(|u_k|^2+|v_k|^2)},\\
    \braket{c_p c_k}&=\delta_{p,-k}\frac{-u_p^* v_k\braket{\eta_{p} \eta_{-k}^\dagger}}{(|u_p|^2+|v_p|^2)(|u_k|^2+|v_k|^2)},\\
    \braket{c_p^\dagger c_k}&=\delta_{p,k}\frac{v_p^* v_k\braket{\eta_{-p} \eta_{-k}^\dagger}}{(|u_p|^2+|v_p|^2)(|u_k|^2+|v_k|^2)},\\
    \braket{c_p c_k^\dagger}&=\delta_{p,k}\frac{u_p^* u_k\braket{\eta_{p} \eta_{k}^\dagger}}{(|u_p|^2+|v_p|^2)(|u_k|^2+|v_k|^2)}.
\end{align}    
\end{subequations}
Performing the Fourier transformation, the real-space correlation functions read as
\begin{subequations}
\begin{align}
    \braket{c_m^\dagger c_n^\dagger}\pm\braket{c_m c_n} &= \frac{1}{N}\sum_{k>0}[\frac{u_k v_k^*\mp v_k u_k^*}{|u_k|^2+|v_k|^2}2\sin(k(n-m))],\\
    \braket{c_m^\dagger c_n}\pm\braket{c_m c_n^\dagger} &= \frac{1}{N}[\sum_{k>0}\frac{|v_k|^2\pm |u_k|^2}{|u_k|^2+|v_k|^2}2\cos(k(m-n))].
\end{align}
\end{subequations}
Therefore, spin-spin correlation functions are given by
\begin{subequations}
    \begin{align}
    \braket{\sigma_l^x \sigma_{l+r}^{x}}
    &=\braket{B_l A_{l+1}B_{l+1}...A_{l+r-1}B_{l+r-1}A_{l+r}},\\
    \braket{\sigma_l^y \sigma_{l+r}^{y}} &= (-1)^r\braket{A_l B_{l+1} A_{l+1}...B_{l+r-1}A_{l+r-1}B_{l+r}},\\
    \braket{\sigma_l^z \sigma_{l+r}^{z}} &= \braket{A_l B_l A_{l+r} B_{l+r}},
\end{align}
\end{subequations}
where $A_j=c_j^\dagger + c_j$ and $B_j = c_j - c_j^\dagger$. The pairs of contractions for $A_n$ and $B_m$ are
\begin{subequations}
\begin{align}
    \braket{A_m A_n}&=\delta_{mn}+\frac{1}{\pi}\int_{0}^{\pi}\frac{u_k v_k^*-u_k^*v_k}{|u_k|^2+|v_k|^2}\sin{(k(n-m))}dk,\\
    \braket{B_m B_n}&=-\delta_{mn}+\frac{1}{\pi}\int_{0}^{\pi} \frac{u_k v_k^*-u_k^*v_k}{|u_k|^2+|v_k|^2}\sin{(k(n-m))}dk,\\
    \braket{B_m A_n}&=\frac{1}{\pi}\int_{0}^{\pi}\frac{u_k v_k^*+u_k^*v_k}{|u_k|^2+|v_k|^2}\sin{(k(n-m))}dk \notag\\
    &-\frac{1}{\pi}\int_{0}^{\pi}\frac{|u_k|^2-|v_k|^2}{|u_k|^2+|v_k|^2}\cos{(k(m-n))}dk,\\
    \braket{A_m B_n}&=\frac{1}{\pi}\int_{0}^{\pi}\frac{u_k v_k^*+u_k^*v_k}{|u_k|^2+|v_k|^2}\sin{(k(n-m))}dk \notag\\
    &+\frac{1}{\pi}\int_{0}^{\pi}\frac{|u_k|^2-|v_k|^2}{|u_k|^2+|v_k|^2}\cos{(k(m-n))}dk,
\end{align}    
\end{subequations}
and the magnetization can be obtained as
\begin{equation}
    \braket{\sigma_z} = \frac{1}{\pi}\int_0^\pi\frac{|v_k|^2-|u_k|^2}{|u_k|^2+|v_k|^2}dk.
\end{equation}
One can then compute the corresponding self-normal defined reduced density matrix using the above correlation functions and the magnetization.

To calculate the biorthogonal expectation values, we follow the same procedure, but we consider $\eta_k$ and $\bar{\eta}_k$. The non-vanishing correlation functions for the ground state take the form,
\begin{subequations}
    \begin{align}
        \braket{c_p^\dagger c_k^\dagger}_{LR} &= -\delta_{p,-k} v_p u_k \braket{\eta_{-p}\bar{\eta}_k}_{LR},\\ 
        \braket{c_p c_k}_{LR} &= -\delta_{p,-k}u_p v_k \braket{\eta_{p}\bar{\eta}_{-k}}_{LR},\\
        \braket{c_p^\dagger c_k}_{LR} &= \delta_{p,k}v_p v_k \braket{\eta_{-p}\bar{\eta}_{-k}}_{LR},\\
        \braket{c_p c_k^\dagger}_{LR} &= \delta_{p,k}u_p u_k \braket{\eta_{p}\bar{\eta}_{k}}_{LR},
    \end{align}
\end{subequations}
where $\braket{\eta_{k}\bar{\eta}_k}_{LR}=1$ and the $\braket{\dots}_{LR}$ denote the exceptional values of biorthogonal quantities.
Taking the Fourier transformation into the real space, we have
\begin{subequations}
    \begin{align}
    \braket{c_m^\dagger c_n^\dagger}_{LR}\pm\braket{c_m c_n}_{LR} &= \frac{1}{N}\sum_{k>0}[2\sin(k(n-m))\notag\\&\times(u_k v_k \mp u_k v_k)],\\
    \braket{c_m^\dagger c_n}_{LR}\pm\braket{c_m c_n^\dagger}_{LR} &= \frac{1}{N}[\sum_{k>0}2\cos(k(m-n))(v_k^2\pm u_k^2)].
\end{align}
\end{subequations}
The pairs of contractions are
\begin{subequations}
\begin{align}
    \braket{A_m A_n}_{LR}&=-\braket{B_m B_n}_{LR}=\delta_{mn},\\
    \braket{A_m B_n}_{LR}&=\frac{1}{\pi}\int_{0}^{\pi}2u_k v_k\sin{(k(n-m))}dk\notag\\
    &-\frac{1}{\pi}\int_{0}^{\pi}(v_k^2-u_k^2)\cos{(k(m-n))}dk,\\
    \braket{B_m A_n}_{LR}&=\frac{1}{\pi}\int_{0}^{\pi}2u_k v_k\sin{(k(n-m))}dk\notag\\
    &+\frac{1}{\pi}\int_{0}^{\pi}(v_k^2-u_k^2)\cos{(k(m-n))}dk,
\end{align}
\end{subequations}
the magnetization can be obtained as,
\begin{equation}
    \braket{\sigma_z}_{LR} = \frac{1}{\pi}\int_0^\pi(v_k^2-u_k^2)dk.
\end{equation}
The phase diagram of the non-Hermitian XY model has been studied in Ref~\cite{PhysRevA.87.012114} and the phase boundary between the $\mathcal{RT}$ symmetric and broken phases is obtained analytically by considering the energy spectrum, which reads
\begin{subequations}
\begin{align}
    h^2-\gamma^2 &= 1, \quad |h|\geq 1\\
    \gamma &=  0, \quad |h|<1.
\end{align}
\end{subequations}
The phase diagram of an infinite-size system represented by the white dotted line in Fig. \ref{fig8}, exhibits two distinct regimes: (i) a hyperbolic boundary for $|h|\geq1$, and (ii) a critical line segment at $\gamma=0$ for $|h|<1$, which corresponds to the Hermitian XX model.

Figure \ref{fig8} shows various entanglement measures as a function of $\gamma$ and $h$ calculated with the self-normal and biorthogonal defined reduced density matrices. The concurrence and negativity show maximum value along the $\mathcal{RT}$ transition (see Fig. \ref{fig8} (a-b)). The introduction of non-Hermiticity enhances quantum entanglement, with the concurrence and negativity reaching their maximum value precisely at the transition point. On the other hand, the mutual information and quantum coherence show extreme derivatives along the transition points, as shown in Fig. \ref{fig8} (c-d). The correlation grows with increasing non-Hermiticity and reaches its maximal enhancement rate precisely at the exceptional point.

Figure \ref{fig8} (e-h) shows the real part of the biorthogonal quantities as a function of $h$ and $\gamma$. Singularities in these quantities are observed at the transition points which outline phase boundaries that quantitatively agree with self-normal measures. Note that the biorthogonal concurrence defined in Eq. (\ref{eq4}) also fails to detect the $\mathcal{RT}$ transition in some parameter regimes in the non-Hermitian XY model (not shown here), which is similar to the non-Hermitian XXZ model. Instead, the unconstrained concurrence in Eq. (\ref{eq:newC}) can successfully determine the phase boundaries (Fig. \ref{fig8}(e)). In addition, the imaginary part of biorthogonal quantities except for the negativity, can also signal the $\mathcal{RT}$ transition in the model. The imaginary part of unconstrained concurrence, mutual information and quantum coherence shows maximum value at the transition points (see Fig. \ref{fig8} (i), (k-l)). The general feature of the phase diagrams is that the second-order QPT in the Hermitian case changes from the Ising transition to the exceptional points as Hermiticity is turned on \cite{Liu_2021}.

\section{Conclusion}\label{sec5}
In summary, we study the non-Hermitian phase transition and the quantum phase transitions under the effect of non-Hermiticity from the perspective of quantum information. We find that self-normal defined entanglement measures and the biorthogonal defined negativity, mutual information and quantum coherence can detect the phase boundaries of $\mathcal{PT}$ ($\mathcal{RT}$) transition, the first-order and the Ising transitions and they give consistent phase boundaries in the models considered. For the BKT transition, we find that the concurrence and negativity give a consistent result at small strength of non-Hermiticity, while other investigated entanglement measures show no signal due to the small system size accessible by the exact diagonalization or the deficiency in the definition itself. In general, we observe that the critical points stemming from the Hermitian case change into the non-Hermitian exceptional points as the non-Hermiticity of the models increases. In particular, in the non-Hermitian XXZ model with a staggered imaginary field, while the first-order and BKT transitions occurring in the Hermitian case survive in a small imaginary field, they merge with the $\mathcal{PT}$ transition as the non-Hermiticity of the model increases. Furthermore, the marriage for the BKT transition with the $\mathcal{PT}$ transition takes place sooner than that for the first-order transition as the non-Hermiticity increases. Similar also occurs in the transverse-field XY model with non-Hermitian spin-spin interactions. The Ising transition point in the Hermitian case becomes the exceptional point as the non-Hermiticity is turned on. 

Moreover, we reveal that the conventional concurrence calculated from the biorthogonal defined reduced density matrix becomes inadequate for detecting the phase transitions in non-Hermitian systems. To this, we propose the unconstrained concurrence and apply it to study the spin models concerned above. The phase transitions, except the BKT one, in these models are well detected by the unconstrained concurrence, and the resultant phase boundaries agree with those obtained from other quantum information measures. 

For future work, it would be intriguing to investigate the phase transitions and the behavior of the quantum information measures in models with other types of non-Hermiticities. The universal properties of the transition, such as the critical exponents, are also worth further investigation using large-system numerical simulation techniques \cite{PhysRevLett.130.100401,PhysRevB.101.235150,PhysRevB.105.205125,PhysRevLett.132.116503,Guo:2022yfi,zhong2024DMRG}.

\begin{acknowledgments}
We thank Yan-Chao Li, Xiaosen Yang, and Xiang Ji for helpful discussions. This work is supported by Research Grants Council of Hong Kong (Grant No. CityU 11318722) and City University of Hong Kong (Grant No. 9610438, 7006018, 9680320).
\end{acknowledgments}

\begin{appendix}
\begin{figure*}[!tb]  \centering\includegraphics[width=.88\textwidth]{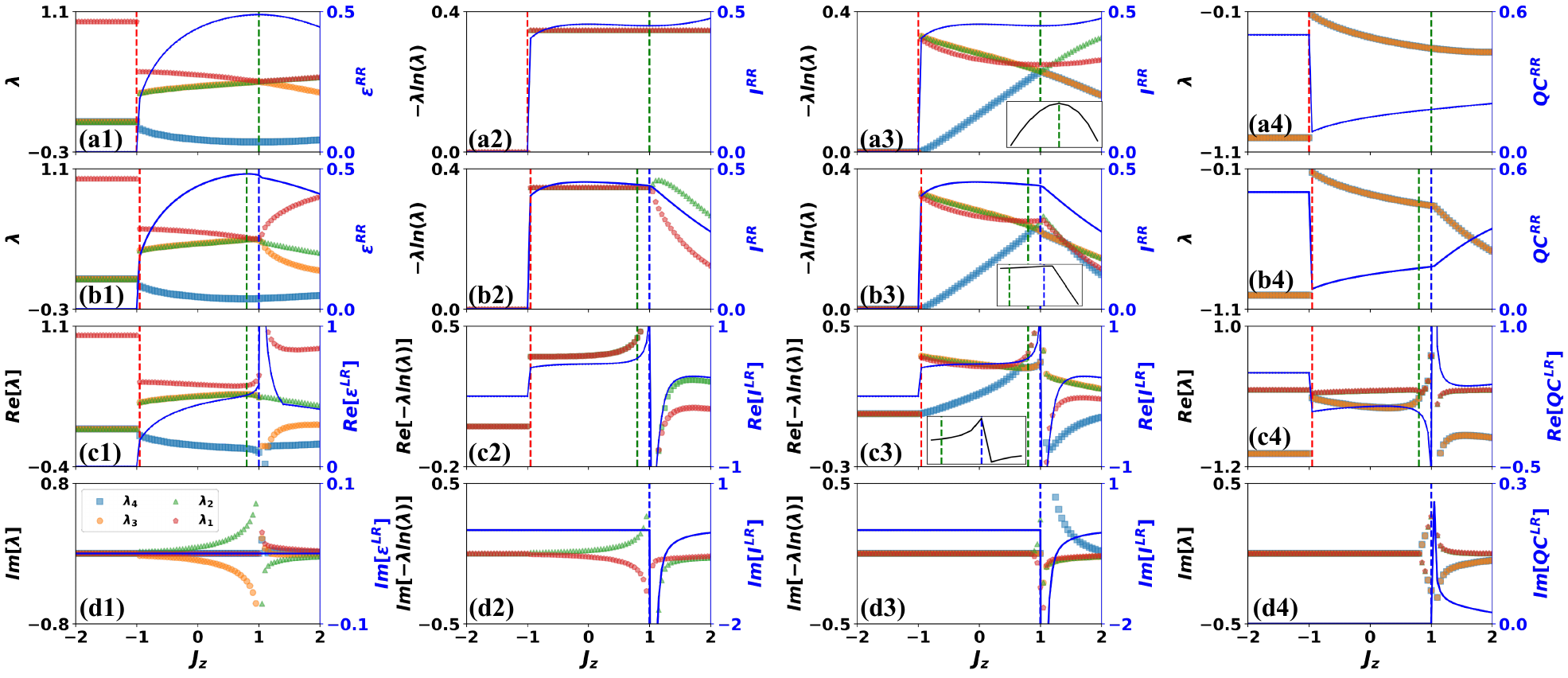}
    \caption{The eigenvalue spectrum of matrices used in different quantum information measures as a function of $J_z$ for $\gamma=0$ (the first row) and $\gamma=0.15$ (the last three rows including self-normal and biorthogonal measures). The red, green, and blue dashed lines are the first-order, BKT, and $\mathcal{PT}$ transitions determined from self-normal concurrence, respectively. The blue solid lines are the corresponding entanglement measures. The first column (a1-d1) is the eigenvalue spectrum of the partially transpose matrix defining the negativity. The middle two columns (a2-d2 and a3-d3) are the entanglement entropy contribution of $\rho_{i}$ and $\rho_{i, j}$, which are related to mutual information and half-chain entanglement entropy. The insets show the entropy $-\sum_i\lambda_i \ln{\lambda_i}$ of $\rho_{ij}$ around the BKT transition ($J_z\in[0.75,1.2]$). The last column (a4-d4) is the eigenvalue spectrum of $[\rho_{i,j},\sigma_{i}^x\otimes I_{j}]^2$, which is related to the quantum coherence.}
    \label{fig9}
\end{figure*} 

\section{Analysis of eigenvalue spectrum used in the quantum information measures}\label{appA}
In this appendix, we investigate the eigenvalue spectrum of matrices that are used to calculate the quantum information measures to reveal the effectiveness in detecting the BKT transition. For negativity, as shown in Fig. \ref{fig9} (a1), we observe that in the Hermitian limit, the negative eigenvalue $\lambda_4$ has a minimum value in the vicinity of the BKT transition, which reflects in the maximum of negativity.  It is also used as a signature in the self-normal negativity in small $\gamma$ as shown in Fig. \ref{fig9} (b1). However, for biorthogonal negativity, due to the singularity around the $\mathcal{PT}$ transition, the signal is concealed, as shown in Fig. \ref{fig9} (c1). For mutual information, as shown in Fig. \ref{fig9} (a2), in the Hermitian limit, the entropy contributions of the single-site reduced density matrix are flat across the BKT transition. On the other hand, in Fig. \ref{fig9} (a3), the entropy contributions of the two-site reduced density matrix have a small peak in the vicinity of the BKT transition. This explains how mutual information and half-chain entanglement entropy signal the BKT transition. However, due to the finite size effect, this signal is weak. When the non-Hermiticity is turned on, it becomes obscured by the singularity due to the $\mathcal{PT}$ transition, as shown in Fig. \ref{fig9} (b3, c3). For quantum coherence, in Fig. \ref{fig9} (a4-d4), the eigenvalue spectrum of matrix $[\rho_{i,j},\sigma_{i}^x\otimes I_{j}]^2$ does not show any signals for the BKT transition. Note that the fluctuations in the imaginary part come from the degeneracy of the real part. Nevertheless, this does not affect the result.
\end{appendix}

% \bibliography{reference}

\end{document}